\documentclass[12pt]{article}

\usepackage{epsfig,latexsym,amsfonts,amsmath,amsthm,amssymb,amsbsy,multirow,slashed,color,hyperref}
\usepackage[font={footnotesize,sl},bf]{caption}

\def\calm         {{\cal M}}

\newsavebox{\uuunit}
\sbox{\uuunit}
    {\setlength{\unitlength}{0.825em}
     \begin{picture}(0.6,0.7)
        \thinlines
        \put(0,0){\line(1,0){0.5}}
        \put(0.15,0){\line(0,1){0.7}}
        \put(0.35,0){\line(0,1){0.8}}
       \multiput(0.3,0.8)(-0.04,-0.02){12}{\rule{0.5pt}{0.5pt}}
     \end {picture}}
\newcommand {\unity}{\mathord{\!\usebox{\uuunit}}}

\def\be{\begin{equation}}
\def\ee{\end{equation}}
\def\bea{\begin{eqnarray}}
\def\eea{\end{eqnarray}}

\newcommand{\eal}[1]{\be \begin{aligned} #1 \end{aligned}\ee}

\newcommand{\beq}{\begin{eqnarray}}
\newcommand{\eeq}{\end{eqnarray}}

\def\a{\alpha}

\def\g{\gamma}

\def\d{\delta}

\newcommand{\ti}[1]{\tilde #1}

\def\to{\rightarrow}

\def\sF{{{ F}\!\!\!\!\hskip.8pt\hbox{\raise1pt\hbox{/}}\,}}
\def\som{{{ \omega}\!\!\!\!\hskip.8pt\hbox{\raise1pt\hbox{/}}\,}}
\def\sJ{{{\rm J}\!\!\!\!\hskip.8pt\hbox{\raise1pt\hbox{/}}\,}}

\newcommand{\En}[1]{\mathop{{\rm E}_{#1(#1)}}}
\newcommand{\SU}{\mathop{\rm SU}}
\newcommand{\SO}{\mathop{\rm SO}}

\newcommand{\USp}{\mathop{\rm {}USp}}

\newcommand{\SL}{\mathop{\rm SL}}
\newcommand{\GL}{\mathop{\rm GL}}

\newcommand{\SLn}[2]{\SL({#1},\mathbb{#2})}

\newcommand{\mb}[1]{\mathbf{#1}}

\renewcommand{\Re}{\operatorname{Re}}
\renewcommand{\Im}{\operatorname{Im}}
\newcommand\e{\mathrm{e}}
\newcommand\iu{\operatorname{i}}
\newcommand\diff{\mathrm{d}}
\newcommand\vol{\operatorname{vol}}

\begin{document}

\begin{titlepage}

\begin{flushright}
{IPhT-T12/047}\\
\end{flushright}
\bigskip
\bigskip
\bigskip
\centering{\Large \bf Black Holes and Fourfolds \vspace{0.9cm}}\\
\bigskip
\bigskip
\centerline{{\bf Iosif Bena,  Hagen Triendl, Bert Vercnocke}}
\bigskip
\centering{Institut de Physique Th\'eorique, \\ CEA-Saclay, CNRS-URA 2306, \\ 
91191 Gif sur Yvette, France\vspace{0.4cm}}
\bigskip
\bigskip
\centerline{{ iosif.bena@cea.fr,~hagen.triendl@cea.fr,~bert.vercnocke@cea.fr} }
\bigskip
\bigskip

\begin{abstract}

We establish the relation between the structure governing supersymmetric and non-supersymmetric four- and five-dimensional black holes and multicenter solutions and Calabi-Yau flux compactifications of M-theory and type IIB string theory. We find that the known BPS and almost-BPS multicenter black hole solutions can be interpreted as GKP compactifications with (2,1) and (0,3) imaginary self-dual flux. We also show that the most general GKP compactification leads to new classes of BPS and non-BPS multicenter solutions. We explore how these solutions fit into $N=2$ truncations, and elucidate how supersymmetry becomes camouflaged.
As a necessary tool in our exploration we show how the fields in the largest $N=2$ truncation fit inside the six-torus compactification of eleven-dimensional supergravity.

\end{abstract}
\end{titlepage}

\setcounter{tocdepth}{2}
\tableofcontents

\section{Introduction}

There is an extensive literature on constructing supersymmetric and non-supersymmetric flux compactifications and a parallel extensive literature on finding supersymmetric and non-supersymmetric multicenter solutions that have the same charges as black holes.
While the physical motivations are different, the technical tools are rather close.
In particular, the equations underlying supersymmetric solutions are well-understood and classified, both on the flux compactification side (see for example \cite{Grana:2000jj,Giddings:2001yu,Grana:2004bg}) in ten dimensions, and on the black hole microstate side for the underlying supergravity in five dimensions \cite{Gauntlett:2002nw,Gutowski:2004yv,Bena:2004de}.
Furthermore, some of the methods for constructing non-supersymmetric solutions from supersymmetric ones are strikingly similar. These methods include slightly deforming the supersymmetric solution by additional fluxes \cite{Grana:2000jj,Giddings:2001yu}, flipping some signs \cite{Goldstein:2008fq}, or writing some effective Lagrangian as a sum of squares for black holes \cite{Ferrara:1997tw,Denef:2000nb,Ceresole:2007wx,Andrianopoli:2007gt,LopesCardoso:2007ky,Janssen:2007rc,Perz:2008kh,Galli:2009bj,Galli:2010mg} or flux backgrounds \cite{Lust:2008zd,Held:2010az}.

A first step towards relating these two research currents has been taken in \cite{Bena:2011pi}, where we found that certain  supersymmetric flux backgrounds of the type \cite{Becker:1996gj} whose ``internal'' manifold contains a hyper-K\"ahler factor are related to non-rotating solutions in the classification of \cite{Gauntlett:2002nw,Gutowski:2004yv,Bena:2004de}. A very intriguing revelation of this relation has been the existence of so-called ``camouflaged supersymmetry'': certain supersymmetric solutions of $N=8$ supergravity are non-supersymmetric in \emph{all} $N=2$ truncations in which they fit. Hence, their supersymmetry is camouflaged in $N=2$ supergravity.

The first purpose of this paper is to deepen the relation between supersymmetric flux compactifications and multicenter solutions found in  \cite{Bena:2011pi}, and to show that this relation extends to non-supersymmetric solutions: some of the non-supersymmetric IIB and M-theory flux compactifications of \cite{Grana:2000jj,Giddings:2001yu,Becker:2001pm} can be reinterpreted as almost-BPS multicenter solutions \cite{Goldstein:2008fq,Bena:2009ev,Bena:2009en,Bossard:2012ge}, while some others give new solutions that lie outside of the almost-BPS class and in general have more dipole charges.

The second purpose of this paper is to blow the cover of camouflaged supersymmetry, by investigating how the field content and supersymmetries of the $T^6$ compactification of M-theory (which gives $N=8$ supergravity) can be truncated to $N=2$. It has been long known that the largest such truncation (with only vector multiplets) is one of the so-called ``Magic Square Supergravities,'' constructed from the Jordan algebra over the quaternions \cite{Gunaydin:1983bi,Gunaydin:1983rk}. Any $N=2$ truncation (without hypermultiplets) of the $N=8$ supergravity should therefore fit into the quaternionic magic supergravity. Therefore, understanding how supersymmetry is camouflaged in the truncations to this supergravity is enough to clarify how this mechanism works in general.

We therefore begin in Section \ref{s:Mtheory} by working out in detail how the fields of eleven-dimensional supergravity compactified on a six-torus project to the multiplets of the $N=2$ quaternionic magic supergravity. This truncation has an $\SU^*(6)$ global symmetry group that is completely determined by a complex structure on $T^6$. The transformation behavior under this complex structure determines the fate of the massless fields coming from the eleven-dimensional metric and three-form potential, and gives the projection of the $N=8$ supersymmetry generators.

We then proceed in Section \ref{s:BH_Flux} to link five-dimensional multicenter solutions to M-theory flux backgrounds on fourfolds that are products of two hyper-K\"ahler spaces. The five-dimensional BPS solutions with a
hyper-K\"ahler base \cite{Gauntlett:2002nw,Gutowski:2004yv,Bena:2004de} correspond to fourfold flux backgrounds with primitive (2,2) flux \cite{Becker:1995kb,Becker:1996gj}, while almost-BPS solutions \cite{Goldstein:2008fq,Bena:2009ev} come from fourfolds with self-dual but supersymmetry-breaking fluxes \cite{Becker:2001pm}. When one of the hyper-K\"ahler spaces is $T^4$ the fourfold flux backgrounds are dual to GKP solutions \cite{Giddings:2001yu}, and the almost-BPS five-dimensional solutions correspond to backgrounds with (0,3) flux \cite{Grana:2005jc}.

However, a few surprises are in store. First, there are supersymmetric flux backgrounds that give solutions that are neither in the five-dimensional BPS \cite{Gauntlett:2002nw,Gutowski:2004yv,Bena:2004de} or almost-BPS classes \cite{Goldstein:2008fq,Bena:2009ev}. Those are the solutions with camouflaged supersymmetry. Second, there exist new non-supersymmetric five-dimensional solutions that have a hyper-K\"ahler base space, and that have many species of both self-dual and anti-self-dual fluxes on this base. The third, and perhaps the most striking result, is that some supersymmetric flux compactifications give supersymmetric solutions of five-dimensional {\em gauged} supergravity with a time-dependent axion (but constant axion field strength). Unlike any other known supersymmetric solutions of five-dimensional gauged supergravity \cite{Gutowski:2004yv,Klemm:2000gh,Klemm:2000nj}, the solutions we find have a hyper-K\"ahler base space!

In Section \ref{s:Examples} we apply our formalism to a flux compactification of M-theory for which one of the hyper-K\"ahler factors in the internal manifold is a $T^4$. We first use the results of Section \ref{s:BH_Flux} to display which fields enter in the BPS, almost-BPS and camouflaged-supersymmetry solutions, and to give the explicit form of the novel BPS gauged supergravity solutions with a time-dependent axion. We then use the results of Section \ref{s:Mtheory} to show how the solutions with camouflaged supersymmetry fit inside $N=2$ truncations of $N=8$ five-dimensional supergravity, and how their Killing spinors do not.

In Section \ref{s:Explicit} we construct a simple explicit example of a solution with camouflaged supersymmetry. The parameters of the general solutions with camouflaged supersymmetry are functions that depend on four variables, and hence these solutions are generically rather complicated. If one assumes that the base hyper-K\"ahler space is a Gibbons-Hawking or a Taub-NUT space, then the solution is completely determined by specifying several harmonic functions on the $\mathbb{R}^3$ base of Taub-NUT. We work out a simple single-center solution in $\mathbb{R}^4$ and in Taub-NUT, and find that it describes a black hole in five dimensions that has a finite horizon area, but whose scalars diverge at the horizon.

\bigskip

\noindent {\bf Outlook}

\bigskip

\noindent Before beginning our investigation it would be useful to remind ourselves what one hopes to obtain by relating the technologies of flux compactifications to the technologies of multicenter black holes and microstates.  The main hope is to apply flux-compactification technology to find new black hole multicenter and microstate solutions with interesting physics, as we have began doing in  \cite{Bena:2011pi}, and to construct new flux compactifications or holographically-useful asymptotically-$AdS$ solutions by using black-hole multicenter technology.

The most promising technology on the flux compactification side is the writing of effective Lagrangians as sums of squares of calibrations \cite{Lust:2008zd,Held:2010az}, which allows one to construct non-supersymmetric backgrounds that depend on functions of several variables by solving first-order equations. This has also been done for multicenter black holes and microstates by finding the first-order system underlying all five-dimensional solutions that admit three kinds of M2 calibrations, or M2 ``floating branes'' \cite{Bena:2009fi}. However, there exist more exotic types of calibrations \cite{Bena:2011ca} that appear to underly general non-extremal solutions like the five-dimensional dipole black ring \cite{Elvang:2004xi} or the so-called JMaRT solution \cite{Jejjala:2005yu}. If one found a way to use these calibrations to obtain a first-order system of equations governing these solutions one would be able to find a simple way of constructing multiple dipole black rings, or multi-bubble JMaRT solutions, which would be very interesting in establishing how the so-called fuzzball proposal\footnote{See \cite{Mathur:2005zp, Bena:2007kg, Mathur:2008nj, Balasubramanian:2008da, Skenderis:2008qn, Chowdhury:2010ct} for reviews.} applies to non-extremal black holes.

From the black hole side, there are two technologies that allow one to construct non-supersymmetric solutions governed by a set of first-order equations. The first is to dualize solutions that have floating branes \cite{Goldstein:2008fq,Bena:2009fi} and to obtain the most general non-BPS solution in their duality orbit \cite{Dall'Agata:2010dy}. The second is to use nilpotent orbits to construct interacting-non-BPS solutions \cite{Bossard:2011kz}. These classes of solutions can have rather unexpected physics\footnote{For example the near-horizon-extremal-Kerr metric can be found \cite{Bena:2012wc} in the infrared of certain solutions obtained in \cite{Dall'Agata:2010dy}}, and if one could extract new flux compactifications from the classes of solutions above, their physics would certainly be interesting.

%%%%%%%%%%%%%%%%%%%%%%%%%%%%%%%%%%%%%%%%%%%%%%%%%%%%%%%%%%%%%%%%%%%%%%%%%%%%%%%%%%%%%%%%%%%%%%%%%%%%%%%%%%
%%%%%%%%%%%%%%%%%%%%%%%%%%%%%%%%%%%%%%%%%%%%%%%%%%%%%%%%%%%%%%%%%%%%%%%%%%%%%%%%%%%%%%%%%%%%%%%%%%%%%%%%%%
\section{M-theory on \texorpdfstring{$T^6$}\ }\label{s:Mtheory}

In this paper we mainly consider M-theory compactified on a six-dimensional torus $T^6$. This leads to a five-dimensional ``maximal'' supergravity, (with $32$ supercharges), which we will denote by $N=8$ supergravity. The purpose of this section is to discuss the ``largest'' truncation of this five-dimensional $N=8$ supergravity to  $N=2$ supergravity (with $8$ supercharges), in terms of the eleven-dimensional fields and the geometry of the internal space.\footnote{The notion ``largest'' here refers to the number of vector multiplets in the $N=2$ theory.}

%%%%%%%%%%%%%%%%%%%%%%%%%%%%%%%%%%%%%%%%%%%%%%%%%%%%%%%%%%%%%%%%%%%%%%%%%%%%%%%%%%%
\subsection{Maximal supergravity in five dimensions}

%%%%%%%%%%%%%%%%%%%%%%%%%%%%%%%%%%%%%%%%%%%%%%%%%%%%%%%%%%%%%%%%%%%%%%%%%%%%%%%%%%%%

We review the reduction to five dimensions of the eleven-dimensional bosonic fields, namely the metric components $G_{MN}$ and the three-form potential $A_3$. We consider a Kaluza-Klein reduction on $T^6$ and keep only the massless modes. The fields of the five-dimensional theory naturally form representations of the $\SLn 6 R$ group of reparameterizations of the six-torus. We expect the reduced theory to have (at least) this $\SLn 6 R$ global symmetry, as well as an $\mathbb{R}$ dilaton shift symmetry related to the scaling of the torus volume or `breathing mode.'
However, because of supersymmetry the fields assemble in representations of an even larger symmetry group, ${\En 6}$, that contains the geometric symmetries $\SLn 6 R \times \mathbb R$ as a subgroup. As a result, we obtain the bosonic field content of five-dimensional $N=8$ supergravity with global symmetry group ${\En 6}$. As we explain below, the 42 scalars of the theory parameterize the coset ${\En 6}/{\USp(8)}$, while the 27 vectors form an irreducible representation of the global symmetry group ${\En 6}$.

We denote by $x^M$ the eleven-dimensional spacetime coordinates, by $x^\mu$ the five-dimensional spacetime directions and by $y^i$ the internal directions. The eleven-dimensional metric $G_{MN}$ decomposes into the five-dimensional metric $G_{\mu\nu}$, six Kaluza-Klein vectors $A^i$ corresponding to the components $G_{\mu i}$ and 21 scalar fields corresponding to the internal metric $G_{ij}$.  The Kaluza-Klein vectors $A^i$ transform in the fundamental representation ${\bf 6}$ of $\SLn 6 R$, the internal components $G_{ij}$ decompose into a ${\bf 20}$ (symmetric traceless) and a singlet $\phi$ (the trace). The latter measures the overall size of the torus and is usually called the dilaton. In a conventional KK ansatz, we would write:
\bea
ds^2_{11} &=& e^{2\alpha \phi} ds_5^2  + e^{2\beta \phi} \calm_{ij} (\diff y^i + A^i)\otimes (\diff y^j + A^j) \ ,
\eea
where $\alpha,\beta$ are constants and $\calm_{ij}=G_{ij}/(\det G_{kl})^{1/6}$ has determinant one and contains the $20$ remaining scalars. The latter form the coset $\SLn 6 R / {\rm SO}(6)$, which describes the deformations of the internal metric.

The three-form potential $A_{MNP}$ gives four different kinds of five-dimensional fields. First, there are $15$ vectors $A_{\mu ij}$ and 20 scalars $A_{ijk}$. Furthermore, there are the components  $A_{\mu \nu \rho}$  and $A_{\mu\nu i}$, which are anti-symmetric three- and two-forms in five dimensions. We wish to consider their proper Hodge dualization to a scalar and a vector potential respectively, because we want a five-dimensional theory with only vector and scalar fields. By choosing notation $\tilde A_6$ for the dual field in eleven dimensions we find $6$ additional vectors $\tilde A_{\mu ijk lm}$ and one more scalar $\tilde A_{ijklmn}$,
giving a total of 15 + 6 vectors and 20 + 1 scalars.\footnote{The $A_3$ equation of motion can be written as $0 = d\star_{11} \diff F_4 + \frac 12 F_4\wedge F_4 \equiv \diff F_7$ and we can locally define $F_7 = \diff \tilde A_6$.} Traditionally, the five-dimensional scalar $\tilde A_{ijklmn}$ arising as the Hodge dual of $A_{\mu \nu \rho}$ is called the axion.

In summary, the five-dimensional vectors $A_\mu^I$ form the ${\bf 27}$ representation of ${\En  6}$ which decomposes as
\begin{equation}\begin{aligned}
 {\bf 27}\ \to & \ {\bf 15} + {\bf 6} + {\bf 6} \ , \\
 A_\mu^I\ = & \ \{A_{\mu ij}, A_{\mu ijk lm}, G_{\mu i}\} \ .
\end{aligned} \end{equation}
under ${\En  6}\to \SLn 6 R$.
Furthermore, the $42$ scalars $\phi^a= \{\calm_{ij}, \phi, A_{ijk}, \tilde A_{ijklmn} \}$ of the theory arrange into the coset ${\En  6}/ {\rm \USp}(8)$.

For later use, we consider the branching of the ${\En 6}$ representations in terms of the subgroup $\SLn 6 R \times \SLn 2 R$. The $\SLn 6 R$, the internal torus symmetry, corresponds to the global symmetry of the coset parameterized by the internal metric components $\calm$. The $\SLn 2 R$ is the global symmetry of the axion-dilaton $\tilde A_{ijklmn} + \iu \e^{-2\phi}$. Those two cosets combine to $\frac{\SLn 6 R}{\SO (6)} \times \frac{\SLn 2 R}{\SO (2)}$ and they can be seen as the submanifold of the scalar geometry ${\En 6}/\USp(8)$ where the scalars $A_{ijk}$ are set to zero.
We find the following picture for the scalar fields
\begin{equation}\begin{aligned}
\frac{\En 6}{\USp(8)}  & \qquad \to &  \frac{\SLn 6 R}{\SO (6)} \times \frac{\SLn 2 R}{\SO (2)}  \\
 & \quad A_{ijk}=0& \\
\phi^a & \qquad \to & \{\calm_{ij}, \tilde A_{ijklmn} + \iu \e^{-2\phi} \} \ .
\end{aligned}\end{equation}
where $\phi^a= \{\calm_{ij}, \phi, A_{ijk}, \tilde A_{ijklmn} \}$.
Under the breaking ${\En 6} \to \SLn 6 R \times \SLn 2 R$ the ${\bf 27}$ representation for the vectors branches as
\be
\mathbf{27}  \to  (\mathbf{6}, \mb 2) + (\mb {15}, \mb 1)\ . \label{eq:Branching_SL6}
\ee

%
%%%%%%%%%%%%%%%%%%%%%%%%%%%%%%%%%%%%%%%%%%%%%%%%%%%%%%%%%%%%%%%%%%%%%%%%%%%%%%%%%%%%
\subsection{The largest \texorpdfstring{$N=2$}~ truncation in five dimensions   and \texorpdfstring{$\SU ^* (6)$}~  supergravity}\label{ss:Truncations}

For many applications, it is important to understand supergravity theories with a lower amount of supersymmetry. Of particular interest are $N=2$ theories,  which arise for instance as Calabi-Yau compactifications of M-theory (to five dimensions) and type II string theory (to four dimensions).
Another way to find $N=2$ supergravities is to truncate a theory with more supersymmetry, for instance the $N=8$ supergravity originating from the M-theory compactification we discussed above. It has been shown that the largest possible consistent $N=2$ truncation of maximal five-dimensional supergravity (without hypermultiplets) is the so-called `magical' supergravity of \cite{Gunaydin:1983bi}, whose construction can be given in terms of the Jordan algebra over the quaternions.
This theory has an $\SU ^* (6)$ global symmetry. There are $15$ vectors filling out the ${\bf 15}$ representation of $\SU^* (6)$ and 14 scalars parametrizing the coset $\SU^* (6) / \USp (6)$.
To our knowledge, the interpretation of the $\SU^*(6)$ theory in terms of the M-theory torus compactification has not been discussed in detail in the literature so far.\footnote{The truncation involves breaking the $N=8$ supermultiplet up into $N=2$ supermultiplets and discarding the $N=2$ {\em spin 3/2} and the {\em spin $\leq$ 1/2} multiplets. The $\SU^*(6)$ theory was shown to be the largest consistent truncation in \cite{Gunaydin:1983rk}. For the  reduction of type II to four dimensional supergravity, see \cite{Sen:1995ff,Gunaydin:2009pk}.} The purpose of this section is to write down explicitly the relation between the $\SU^*(6)$ fields in five dimensions and the parent eleven-dimensional fields, and to close this gap. Anticipating a bit, we will  see that the truncation to $N=2$ corresponds to a choice of complex structure on the $T^6$.

\subsubsection{Maximal $N=2$ truncations and $\SU ^* (6)$}

As the reader might not be very familiar with $\SU ^* (6)$, a certain real form of $\SLn 6 R$, we first give some more details on this group. $\SU^*(2n)$ consists of those complex matrices $M\in \GL(2n,\mathbb{C})$, commuting with the operator $I K$:
\be
I K M = M I K\,,\label{eq:SU*6_Definition}
\ee
where $K$ is the complex conjugation operator and  $I^2 = -\unity$. From this operational definition, it is clear that $I$ has the interpretation of giving a complex structure to $\mathbb{R}^{2n}$. If we perform a change of basis bringing $I$ to the canonical form $I = \unity_{n\times n} \otimes \iu \sigma_2$, where $\sigma_2$ is the second Pauli matrix, then an element of the Lie algebra $m \in \mathfrak{su}^*{(2n)}$ has the form:\footnote{We define Lie algebra elements as $M = \exp (m)$, without an `$\iu$' in the exponent.}
\be
m = \begin{pmatrix} A & B \\ -B^* & A^*\end{pmatrix}\label{eq:SU*Lie}
\ee
where $B \in \mathfrak{gl}(n,\mathbb{C})$ and $A\in \mathfrak{sl}(n,\mathbb{C})$. See for example \cite{Gilmore:2008zz} for more details. 

Let us return to the truncation of maximal supergravity. Similar to the breaking to the geometric subgroup, the vectors in the representation ${\bf 27}$ of the $N=8$ symmetry group ${\En 6}$ decompose in representations of the subgroup $\SU^*(6) \times \SU(2)$ as:
\be
{\En 6} \to \SU{}^* (6) \times \SU(2):\qquad \mathbf{27} \to  (\mathbf{6}, \mb 2) + (\mb {15}, \mb 1)\label{eq:Branching_SU6} \ .
\ee
Similarly, the $42$ scalars of the coset ${\En  6}/\USp(8)$ split into $14$ vector multiplet scalars parameterizing $\SU^*(6)/\USp(6)$ and $28$ hypermultiplet scalars parameterizing $F_4/(\USp(6)\times \SU(2))$. These two submanifolds of ${\En  6}/\USp(8)$ are not compatible with each other as $\SU^*(6)$ and $F_4$ do not commute as subgroups of ${\En  6}$, and this indicates that they are not decoupled. If we want to keep all vector multiplets in the $N=2$ truncation, all hypermultiplets are truncated away. In the following we consider only the vector multiplets.

We see that only the ${\bf 15}$ in \protect\eqref{eq:Branching_SU6} can survive the truncation to $N=2$, as only these are accompanied by the $14$ scalars.\footnote{Note that one vector, the graviphoton, sits in the gravity multiplet and does not have a scalar in its multiplet. Therefore, the number of scalars is one lower than the number of vectors.} In the next paragraph, we identify the ${\bf 15}$ vectors and the 14 scalars of the maximal $N=2$ truncation in terms of the $T^6$ geometry.

\subsubsection{Internal geometry}\label{ss:SL3C}

The choice of complex structure  on the torus is crucial in treating the geometric interpretation of $\SU^*(6)$ supergravity. The choice of complex structure corresponds to choosing  complex coordinates $z^\alpha, \bar z ^{\bar \beta}$ in terms of a set of real coordinates $y^i$, or more precisely, of choosing the embedding of an $\SLn 3 C$ subgroup in the $\SLn 6 R$ group of torus reparameterizations. The choice of complex structure fixes  a preferred matrix $I$ and through \eqref{eq:SU*6_Definition} it singles out an $\SU^* (6)$ subgroup of $E_{(6)6}$.

Let us now understand the decomposition for the vectors in \eqref{eq:Branching_SU6} and for the scalars into $\SU^*(6)/\USp(6)$ and $F_4/(\USp(6)\times \SU(2))$ by considering the largest common subgroup of $\SLn 6 R \times \SLn 2 R$ and $\SU^*(6)\times \SU (2)$.\footnote{We thank Sergio Ferrara very much for proposing this strategy.} This group is $\SLn 3 C \times U(1) \times U(1)$ as can be seen in the following way. Consider first the $\SU^*(2n)$ Lie algebra element $m$ in \eqref{eq:SU*Lie} and choose coordinates such that $I$ has the canonical form. For $m$ to be an element of the Lie algebra $\mathfrak{sl}(2n,\mathbb{R})$ as well, it must be a real and traceless matrix. This restricts its elements:
\be
m = A \otimes \unity_{2\times 2}+ B \otimes \iu\sigma_2 \label{eq:SU*LieSL}\,,\qquad A\in \mathfrak{sl}(n,\mathbb{R})\,,  B \in \mathfrak{gl}(n,\mathbb{R})\ .
\ee
We conclude that the largest common subgroup is\footnote{This latter isomorphism can be made concrete by considering the mapping:
$\unity_{2\times 2}\leftrightarrow 1\,, \sigma \leftrightarrow i$, such that the Lie algebra element is mapped into $m \to \{A + i \tilde B \in \mathfrak{sl}(n,\mathbb{C})\,,{\rm Tr}\, iB\in\mathfrak{u}(1) \equiv \mathfrak{so}(2)\}$ where $\tilde B$ denotes the traceless part of the matrix $B$.}
\be
SL(n,\mathbb{R}) \,\cap\,SU^*(2n)
\cong SL(n,\mathbb{C}) \times U(1) \ .
\ee
The $U(1)$ is generated by $(\unity \otimes \iu \sigma_2)$ and corresponds in our case (for $n=3$) to $I$, the complex structure on $T^6$. The group $\SLn 3 C$ are the reparametrizations that do no affect the complex structure.
Furthermore, the common subgroup of the factors $SL(2,\mathbb{R})$ and $\SU(2)$ is given by
\be
SL(2,\mathbb{R}) \,\cap\, \SU(2)
\cong U(1) \ .
\ee
We arrive at the full breaking pattern of ${\En  6}$ into its subgroups in Table \ref{tab:group_breaking}.
Similarly, one can work out the breaking of the maximal compact subgroup $\USp(8)$, as given in Table \ref{tab:group_breaking_comp}.
\begin{table}[ht!]
\centering
\begin{tabular}{|ccc|}
\hline
${\En  6}$ & $\to$ & $\SLn 6 R \times \SLn 2 R$ \\
$\downarrow$ && $\downarrow$ \\
$\SU^*(6)\times \SU(2)$ & $\to$ & $\SLn 3 C \times U(1) \times U(1)$ \\
\hline
\end{tabular}
\caption{
The breaking of ${\En  6}$ into its subgroups. In vertical direction the breaking to $N=2$ is displayed, in horizontal direction the breaking to the geometric subgroup.
\label{tab:group_breaking}}
\end{table}
\begin{table}[ht!]
\centering
\begin{tabular}{|ccc|}
\hline
$\USp(8)$ & $\to$ & $\SU(4) \times U(1)$ \\
$\downarrow$  && $\downarrow$ \\
$\USp(6)\times \SU(2)$ & $\to$ & $\SU(3) \times U(1) \times U(1)$ \\
\hline
\end{tabular}
\caption{
The breaking of $\USp(8)$ into its subgroups. In vertical direction the breaking to $N=2$ is displayed, in horizontal direction the breaking to the geometric subgroup.
\label{tab:group_breaking_comp}}
\end{table}

With the data of Table \ref{tab:group_breaking} and \ref{tab:group_breaking_comp} we are now able to determine how the components of the ten-dimensional fields arrange into representations of $\SU^*(6) \times \SU(2)$.  We can identify the field content of the $\SU^*(6)\times \SU(2)$ representations by using their decompositions under the `geometric' subgroup $\SLn 3 C \times U(1) \times U(1)$. First we decompose the gauge fields and scalars that we got from the dimensional reduction into representations of $\SLn 3 C$ using that the torus one-forms split into holomorphic ($\mb 3$ of $\SLn 3 C$) and antiholomorphic ($\bar{\mb 3}$ of $\SLn 3 C$) one-forms. The result of the decomposition can be found in Table \ref{tab:Branching_SL3C}.
\begin{table}[ht!]
\centering
\begin{tabular}{cll||ccc}
\multicolumn{3}{c||}{Field content:}&$\SLn 6 R$ & $\SLn 3 C \times U(1)_{\rho_6}$&\\
\hline
\hline
Vectors: &$A_{\mu ij}$ &$=\{A_{\mu \a \bar \beta}, A_{\mu \bar \a \bar \beta}, A_{\mu \a \beta}\}$&$\mb{ 15}$& $(\mb 8_{\bf 0} + \mb 1_{\bf 0}) + \mb 3_{\bf -2} + \bar{\mb 3}_{\bf +2}$&\\
&$G_{\mu i} $ &$= \{G_{\mu \a},G_{\mu \bar \a}\}$&$\mb 6$&$\mb 3_{\bf +1} + \bar{\mb 3}_{\bf -1}$\\
&$\tilde A_{\mu ijklm} $ &$= \{\tilde A_{\mu \bar \a \bar \beta \g \d \epsilon},\tilde A_{\mu \a\beta \bar\g \bar\d \bar\epsilon}\}$ &${\mb 6}$&$\mb 3_{\bf +1} + \bar{\mb 3}_{\bf -1}$\\
\hline
\hline
Scalars:&
$A_{ijk}$ &$=\{A_{\a \beta \g},A_{\bar \a \bar \beta \bar \g}, A_{\bar \a \beta \gamma}, A_{ \a \bar\beta\bar \gamma} \} $&$\mb {20}$&$\mb 1_{\bf +3} + \mb 1_{\bf -3}+ (\mb 3_{\bf +1} + \bar{\mb 6}_{\bf +1})$\\
&&&& $ + (\bar {\mb 3}_{\bf -1} + \mb 6_{\bf -1})$\\
&$G_{ij}$ &$= \{G_{\bar \a \bar \beta}, G_{\a \beta}, G_{\a \bar \beta}\}$&$\mb {20'} + \mb 1$&$\bar{\mb 6}_{\bf -2} + \mb 6_{\bf +2} + (\mb 8_{\bf 0} + \mb 1_{\bf 0})$\\
 &$\tilde A_{ijklmn}$&$=A_{\alpha \beta \gamma \bar \delta \bar \epsilon \bar \sigma} $ &$\mb{1}$ & $\mb{1}_{\bf 0}$ \\
\end{tabular}
\caption{Higher-dimensional origin of the five-dimensional vectors and scalars, their $\SLn 6 R$ representations and branching under the $\SLn 3 C \times U(1)$ subgroup. The $\mb 3, \bar{\mb 3}$ and the $\mb 6,\bar{\mb 6}$ appearing in the vector sector are the irreducible (anti-)symmetric parts of the two-tensor representations.
\label{tab:Branching_SL3C}}
\end{table}

It is then crucial to understand the relations between the $U(1)$'s coming from $\SU^*(6)\times \SU(2)$ and from $\SLn 6 R \times \SLn 2 R$, cf.\ Table \ref{tab:group_breaking} and \ref{tab:group_breaking_comp}. Let us denote the generator for the $U(1)$ inside $\SU^*(6)$ by $u_6$ and the one coming from $\SU(2)$ by $u_2$. Similarly, $\rho_2$ and $\rho_6$ denote the $U(1)$'s coming from $\SLn 2 R$ and $\SLn 6 R$. Then we can derive from Table \ref{tab:group_breaking_comp} that
\begin{equation}\label{eq:U1charges}\begin{aligned}
 u_6 =& \tfrac12 (3\rho_2 + \rho_6) \ , &\qquad  & \rho_6 = \tfrac12 (u_6-3u_2)  \ , \\
 u_2 =& \tfrac12 (\rho_2 - \rho_6) \ , & \qquad & \rho_2 = \tfrac12 (u_6+u_2) \ .
\end{aligned} \end{equation}
where the overall signs are just conventional.

We start with the vector multiplets, where the breaking is given by \eqref{eq:Branching_SU6}.
In terms of $\SLn 3 C \times U(1)_{u_6} \times U(1)_{u_2}$ representations, this becomes
\begin{equation}\begin{aligned}
(\mb {15}, \mb 1) \to & \mb 8_{{\bf 0},{\bf 0}} \oplus \mb 1_{{\bf 0},{\bf 0}} \oplus \mb 3_{{\bf +2},{\bf 0}} \oplus \bar{\mb 3}_{{\bf -2},{\bf 0}} \ ,\\
(\mathbf{6}, \mb 2) \to & \bar{\mb 3}_{{\bf +1},{\bf +1}}\oplus {\mb 3}_{{\bf -1},{\bf +1}} \oplus \bar{\mb 3}_{{\bf +1},{\bf -1}}\oplus {\mb 3}_{{\bf -1},{\bf -1}} \ .
\end{aligned}\end{equation}
From \eqref{eq:U1charges} we see that the representations under $\SLn 3 C \times U(1)_{\rho_6} \times U(1)_{\rho_2}$ are
\begin{equation}\begin{aligned}
(\mb {15}, \mb 1) \to & \mb 8_{{\bf 0},{\bf 0}} \oplus \mb 1_{{\bf 0},{\bf 0}} \oplus \mb 3_{{\bf +1},{\bf +1}} \oplus \bar{\mb 3}_{{\bf -1},{\bf -1}} \ ,\\
(\mathbf{6}, \mb 2) \to & \bar{\mb 3}_{{\bf -1},{\bf +1}}\oplus {\mb 3}_{{\bf -2},{\bf 0}} \oplus \bar{\mb 3}_{{\bf +2},{\bf +0}}\oplus {\mb 3}_{{\bf +1},{\bf -1}} \ .
\end{aligned}\end{equation}
The representations on the right-hand side can be identified with the representations appearing in Table \ref{tab:Branching_SL3C}. For instance, the representation $\mb 3_{{\bf +1},{\bf +1}}$ is given by $G_{\mu \alpha} + \iu (\ast_6 \tilde A_\mu)_\alpha$, while the ${\mb 3}_{{\bf +1},{\bf -1}}$ is given by $G_{\mu \alpha} - \iu (\ast_6 \tilde A_\mu)_\alpha$.
Therefore we identify the components of the $({\bf 15},{\bf 1})$ representation of $\SU^*(6)\times \SU(2)$ as $\{A_{\mu ij} + I^k_i I^l_j A_{\mu kl}, G_{\mu i} - I_i^j(\ast_6 \tilde A_\mu)_j \}$. These are the vectors that remain in the $N=2$ supergravity, while the others are projected out. The result is displayed in Table \ref{tab:Branching_SU6_vect}.

\begin{table}[ht!]
\centering
\begin{tabular}{c||ccc}
&$\SU^* (6)$ & &${\En 6}$\\[2mm]
\hline
&&&\\[-12pt]
$A_{\mu ij} + I^k_i I^l_j A_{\mu kl},\ G_{\mu i} - I_i^j(\ast_6 \tilde A_\mu)_j$&$\mb{15}$&\multirow{3}{*}{$\left.\begin{array}{c}\\ \\ \end{array}\right\}$}&\\
$G_{\mu i} + I_i^j(\ast_6 \tilde A_\mu)_j$&$\mb 6$& &$\mb{27}$\\
$A_{\mu ij} - I^k_i I^l_j A_{\mu kl}$ & $\mb 6$&&
\end{tabular}
\caption{Vectors in five-dimensional supergravity with their higher-dimensional origin (first column) and representations under $\SU^*( 6)$ and $\En 6$. \label{tab:Branching_SU6_vect}}
\end{table}

Now let us turn to the scalars. In $N=8$ supergravity, the scalars form the coset ${\En  6}/\USp(8)$ which corresponds to the ${\bf 42}$ representation of $\USp(8)$. The $N=2$ truncation breaks $\USp(8) \to \USp(6) \times \SU(2)$. The scalars split accordingly into
\begin{equation}
 {\bf 42} \to ({\bf 14},{\bf 1}) \oplus ({\bf 14'},{\bf 2}) \ .
\end{equation}
The first term gives the vector scalars that survive the $N=2$ projection and form the coset $\SU^*(6)/\USp(6)$. The second term would be hyper scalars but they do not survive the projection.
In terms of the geometric subgroup $\SLn 3 C \times U(1)$, the ${\bf 14}$ breaks into
\begin{equation}
 {\bf 14} \to {\bf 8}_{\bf 0} \oplus {\bf 3}_{\bf +2} \oplus {\bf \bar 3}_{\bf -2} \ .
\end{equation}
These representations are identified with the scalars $\{ \calm_{\a \bar \beta},  A_{\alpha \gamma \bar \gamma}\delta^{\gamma \bar \gamma},  A_{\bar \alpha \bar \gamma \gamma }\delta^{\bar \gamma \gamma} \}$, as can be read off from Table \ref{tab:Branching_SL3C}. All other scalars are projected out of the $N=2$ theory.
Note that only those components of the internal metric that preserve the complex structure $I$ survive.

\subsection{Spinors}\label{ss:SpinorTruncations}

We now study the effects of the $N=2$ truncation on the eleven-dimensional Killing spinors. In $N=8$ the internal components of the Killing spinors transform under $\USp(8)$ in the fundamental representation given by
\begin{equation} \label{eq:fund_USp}
\eta = \left( \begin{aligned} \eta_+ \\ {\cal C} \eta_- \end{aligned} \right) \ ,
\end{equation}
where $\eta_\pm$ are the chiral components of the internal spinor and ${\cal C}$ is the six-dimensional charge conjugation matrix.

For this, we consider the breaking of the R-symmetry group $\USp(8)$ of the $N=8$ theory in five dimensions to the $N=2$-supersymmetric truncation. Following Table \ref{tab:group_breaking_comp}, it is broken as
\begin{equation}
\label{eq:R_sym_breaking}
\USp(8) \to \USp(6) \times \SU(2)\ .
\end{equation}
This breaking of the R-symmetry group corresponds to a projection on the space of internal Killing spinors.
The $N=2$ Killing spinors are singlets under the action of the $\USp(6)$ factor and the $\SU(2)$ factor comprises the $N=2$ R-symmetry. This means that the action of the $\USp(6)$ generators vanishes on the $N=2$ spinors. This gives us the projection operators for mapping the $N=8$ spinors to the $N=2$ subspace. In general it is sufficient to consider only a set of Cartan generators $g_i,\ i=1,2,3,$ of $\USp(6)$ as projection operators. Then the projection onto the $N=2$ spinors is given by
\begin{equation}
g_i \eta = 0 \ , \quad i=1,2,3 \ .
\end{equation}

In order to find a convenient set of Cartan generators $g_i$ of $\USp(6)$, we consider the geometric subgroup $Spin(6)\equiv \SU(4)$ of the R-symmetry group, which acts on the components $\eta_\pm$ in \eqref{eq:fund_USp} separately.
It breaks accordingly as
\begin{equation}
Spin(6) \equiv \SU(4) \to \SU(3) \times U(1) \ ,
\end{equation}
cf.\ Table \ref{tab:group_breaking_comp}. In Section \ref{ss:SL3C}, we denoted the generator of the $U(1)$ factor as $\rho_6$. On the spinors it just acts as $\rho_6 = \slashed J$, where $J$ is the K\"ahler two-form related to $I$. Furthermore, for the Cartan generators of $\SU(3)$ we can choose any two elements of $Spin(6)$ that commute with each other and with $\rho_6$. Since $\SU(3)$ is a subgroup of $\USp(6)$, these two Cartan generators give appropriate $g_1$ and $g_2$. A generator of $\USp(6)$ that is obviously commuting with $\SU(3)$ is $u_6$. Therefore, $g_3=u_6$ is our third Cartan generator of $\USp(6)$. Utilizing \eqref{eq:U1charges} we see that $g_3$ is given by $\tfrac12 (3\rho_2 + \rho_6)$. Here, $\rho_2$ acts as the matrix
\begin{equation}
 \rho_2 = \left( \begin{aligned} 0 && 1 \\ -1 && 0 \end{aligned} \right) \otimes 1
\end{equation}
on \eqref{eq:fund_USp}.

To visualize these conditions, we specify a complex coordinate system $z^1,z^2,z^3$ on $T^6$. The K\"ahler form on $T^6$ has the form
\begin{equation}
J = \sum_i \tfrac {\iu} 2 \diff z^i \wedge \diff \bar z^i\ .
\end{equation}
With real coordinates as in Section \ref{s:Examples} ($z^1 = y^5 + \iu y^8,\ z^2 = y^6 + \iu y^7,\ z^3 = x^9 + \iu x^{10}$), we then have
\begin{equation}
\slashed J = \Gamma^{85} + \Gamma^{67} + \Gamma^{9,10}\ .
\end{equation}
We find that  $g_1 = \Gamma^{85}-\Gamma^{67}$ and $g_2 = \Gamma^{67}-\Gamma^{9,10}$ both commute with $\slashed J$ and hence they span the $\SU(3)$ Cartan algebra. This gives us the projection conditions:
\begin{equation}
(1 -\Gamma^{5678})\ \eta_\pm = 0\ , \qquad (1 -\Gamma^{679,10})\ \eta_\pm =0\ , \qquad (\Gamma^{85} + \Gamma^{67} + \Gamma^{9,10}) \eta_\pm +3 {\cal C} \eta_\mp = 0 \ .\label{eq:SpinorsProjection}
\end{equation}

%%%%%%%%%%%%%%%%%%%%%%%%%%%%%%%%%%%%%%%%%%%%%%%%%%%%%%%%%%%%%%%%%%%%%%%%%%%%%%%%%%%%%%%%%%%%%%%%%%%%%%%%%%
%%%%%%%%%%%%%%%%%%%%%%%%%%%%%%%%%%%%%%%%%%%%%%%%%%%%%%%%%%%%%%%%%%%%%%%%%%%%%%%%%%%%%%%%%%%%%%%%%%%%%%%%%%
\section{Black holes and flux compactifications}\label{s:BH_Flux}

In this section we investigate the relation between black hole solutions and flux compactifications. On the black hole side, we wish to consider compactifications of string/M-theory to five dimensional solutions that can describe black holes, black rings and their microstate geometries. In M-theory, such geometries are of the type
\be
{\cal M}_{1,10} = \mathbb{R}_t \times M_4 \times \tilde M_6\,.
\ee
The geometry depends on the four-dimensional base space $M_4$; the compact space $\tilde M_6$ and the time direction are non-trivially fibered over $M_4$. For supersymmetric solutions, $M_4$ is hyper-K\"ahler and $\tilde M_6$  is Calabi-Yau.

We can interpret this geometry as a flux compactification to Minkowski space in $d=1$ (only a time direction), where $M_4 \times \tilde M_6$ is the internal space and the black hole redshift factor ($g_{tt}$ metric component) acts as the warp factor. Of course $M_4$ describes the four-dimensional space of the black hole spacetime and is non-compact. However, all techniques in the flux compactification literature are still largely applicable, as they mostly concern local properties of the solution, such as supersymmetry and solving the equations of motion.

From this crude picture we see that a black hole geometry can be interpreted as a one-dimensional flux vacuum. There is however no real technical gain in the study of black hole solutions from this analogy, as for obvious reasons one-dimensional flux vacua have not received much attention in the literature.\footnote{A first attempt to understand such backgrounds better can be found in \cite{Tomasiello:2011eb}.} Flux compactifications of M-theory to three dimensions on the other hand have been studied in great detail, starting with the class of Minkowski vacuum solutions of \cite{Becker:1996gj,Becker:2001pm}. Therefore we keep to black hole solutions that have a three-dimensional Poincar\'e invariance and fall into this class. In particular this happens when $\tilde M_6$ factorizes as $\ti M_6 = \ti M_4 \times T^2$ and we demand (local) Poincar\'e invariance in the three-dimensional space $\mathbb{R}_t \times T^2$. In this section, we focus on black hole solutions of this type, which can be interpreted as a flux compactification with geometry
\begin{equation}
\label{CY4comp}
{\cal M}_{1,10} = \mathbb{R}_t \times T^2 \times (M_4 \times \tilde M_4) \ .
\end{equation}
A special situation is $\tilde M_4 = T^4$, which we discuss in more detail in Section \ref{s:Examples}.

In Section \ref{ss:Flux}, we review flux compactification of M-theory to three dimensions with internal space $X_8 = M_4 \times \tilde M_4$, where $M_4$ and $\tilde M_4$ are both hyper-K\"ahler.\footnote{Since $\tilde M_4$ is compact, it is therefore either K3 or $T^4$.} This is a special example of a  ``compactification'' on a Calabi-Yau fourfold $X_8$ as discussed in \cite{Becker:1996gj,Becker:2001pm}. We pay special attention to the supersymmetry conditions in terms of fluxes. We also discuss the relation to the GKP class of Minkowski vacua \cite{Giddings:2001yu}.
In Section \ref{ss:BlackHole}, we review black hole, black ring and their microstate solutions in the STU truncation of M-theory. We discuss both the supersymmetric solutions and the `almost-BPS' ones, and show under which conditions they have an interpretation as three-dimensional flux vacuua in M-theory.

%%%%%%%%%%%%%%%%%%%%%%%%%%%%%%%%%%%%%%%%%%%%%%%%%%%%%%%%%%%%%%%%%%%%%%%%%%%%%%%%%%%%
\subsection{Flux compactifications on hyper-K\"ahler spaces}
\label{ss:Flux}

\subsubsection{Calabi-Yau fourfolds in M-theory}

We discuss flux backgrounds of M-theory on a Calabi-Yau fourfold $X_8$ times three-dimensional flat spacetime.
In the absence of flux (and D-branes/orientifold planes) these backgrounds preserve four supercharges. For backgrounds of the form \eqref{CY4comp} the number of preserved supercharges is at least eight, as $X_8=M_4 \times \tilde M_4$ has more Killing spinors.\footnote{For $\tilde M_4=T^4$ there are sixteen unbroken supercharges.} A natural scenario to (partially or completely) break the remaining supersymmetry for such backgrounds by non-trivial fluxes has been established in \cite{Becker:1996gj,Becker:2001pm}. In this setup, there is the possibility of a warp factor induced by an electric flux in the background. The solution is
\begin{equation}\label{eq:general_solution}
\begin{aligned}
 \diff s_{11}^2 = \; \e^{-2A}\diff s_{1,2}^2 + \e^{A} \diff s^2_{CY}(X_8)\ , \\
 G_4 = \diff ( \e^{-3A} \vol_3) + G_4^{\rm mag} \ .
\end{aligned}
\end{equation}
The equations of motion dictate that the internal four-form flux $G^{\rm mag}_4$ on $X_8$ is self-dual. A generic self-dual $G^{\rm mag}_4$ breaks all supersymmetries. However, when the four-form flux on $X_8$ is of cohomology type $(2,2)$ and primitive (orthogonal to the K\"ahler form of $X_8$), the background still preserves 1/2 of the supercharges.

Let us try to understand this in more detail. The most general self-dual flux on $X_8=M_4 \times \tilde M_4$ is
\begin{equation}
 G^{\rm mag}_4 = f ( \vol_4(M_4) + \vol_4(\tilde M_4)) +\sum_{a=1}^{3} \sum_{b=1}^{3} f^{a b} J_a\wedge \tilde J_{b} + \sum_{\alpha=1}^{N}\sum_{\tilde \alpha=1}^{\tilde N} f^{\alpha \tilde \alpha} L_{\alpha} \wedge \tilde L_{\tilde \alpha}\ ,\label{eq:G4_Ansatz}
\end{equation}
where $\vol_4$ denotes the volume forms of each hyper-K\"ahler space, $J_a$ ($\tilde J_{\tilde a}$) are a basis of self-dual two-forms on the hyper-K\"ahler space $M_4$($\tilde M_4$). Furthermore, $N$ ($\tilde N$) denotes the number of the anti-self-dual two-cycles $L_{\alpha}$ ($\tilde L_{\tilde \alpha}$) on $M_4$($\tilde M_4$), which is $(b_2-3)$ in terms of the second Betti number of $M_4$($\tilde M_4$). The coefficients $f, f^{a \tilde a}, f^{\alpha \tilde \alpha}$ parameterize the flux turned on.

Now let us discuss the supersymmetry conditions.
The hyper-K\"ahler manifolds $M_4$, $\tilde M_4$ each have three self-dual two-forms $J_a$,  $\tilde J_a$, $a=1,2,3$, related to the metric by the triplets of complex structures $I_a$ and $\tilde I_a$, cf.\ Appendix \ref{app:HyperKahler}.
The manifold $X_8=M_4 \times \tilde M_4$ admits an $S^3 \times S^3$ family of Calabi-Yau structures
\begin{equation}
J = \rho^a J_a + \tilde \rho^b \tilde J_b \ , \qquad
\Omega = c^a \tilde c^b J_a \wedge \tilde J_b \ ,
\end{equation}
where $\rho^a$ and $\tilde \rho^a$ are real and the $c^a$ and $\tilde c^a$ are complex vectors obeying the conditions
\begin{equation}
\label{eq:condition_ca}
\rho^a c^a = \tilde \rho^a \tilde c^a = 0 \ , \quad
c^a c^a = \tilde c^a \tilde c^a = 0 \ .
\end{equation}
In order for $G_4^{\rm mag}$ to be supersymmetric with respect to $(J,\Omega)$, we must have
\begin{equation}
J \wedge G_4^{\rm mag} = 0 \ , \qquad
\Omega \wedge G_4^{\rm mag} = 0 \ .
\end{equation}
Comparing this with \eqref{eq:G4_Ansatz}, this means for $G_4^{\rm mag}$ to be of the form
\begin{equation}
 G_4^{\rm BPS} = A \left( \vol_4(M_4) + \vol_4(\tilde M_4) - \rho^a {\tilde \rho}^b J_a \wedge \tilde J_b \right) +  \Re(B \, \bar c^a \, \tilde c^b)  J_a\wedge \tilde J_b + f^{\alpha \tilde \alpha} L_{\alpha} \wedge \tilde L_{\tilde \alpha}\ ,\label{eq:G4_susy}
\end{equation}
where $A$ is real and $B$ is complex. This gives the general supersymmetric solution for the background \eqref{CY4comp}. Note that there is no restriction on  the anti-self-dual two-forms ($f^{\alpha \tilde \alpha}$ is unconstrained). The flux $G_4^{\rm BPS}$ is primitive $(2,2)$ with respect to the complex structure
\begin{equation}
 I \equiv \left( \begin{aligned} \hat \rho^a I_a && 0 \\ 0 && \hat  {\tilde \rho}^a \tilde I_a \end{aligned} \right) \ ,\label{eq:ComplexStructure_I}
\end{equation}
where the hat indicates that the vector is normalized to one.

Let us now concentrate on when $\tilde M_4 = T^4$. We can choose
\begin{equation}
J_{T^4} = \tilde \rho^a \tilde J_a = \tfrac \iu  2\diff z^1 \wedge \diff \bar z^1+ \tfrac \iu  2\diff z^2 \wedge \diff \bar z^2\ ,
\end{equation}
and
\begin{equation}
\Omega_{T^4} = \tilde c^a J_a = \diff z^1 \wedge \diff z^2 \ ,
\end{equation}
where $\diff z^1$ and $\diff z^2$ are $\tilde I_3$-holomorphic one-forms.
Inserting this into \eqref{eq:G4_susy} gives
\begin{equation}
\label{eq:G4_susy_T4_f0}
\begin{aligned}
 G_4^{\rm BPS} = & f_0 \left( \vol_4(M_4) + \tfrac12 \diff z^1 \wedge \diff z^2 \wedge \diff \bar z^{1} \wedge \diff \bar z^{2}- \tfrac \iu  2\rho^a  J_a \wedge (\diff z^1 \wedge \diff \bar z^{1} +\diff z^2 \wedge \diff \bar z^{2}) \right)\\ & +  \Re(C \, \bar c^a J_a\wedge \diff z^1 \wedge \diff z^2 )  + \iu f^{\alpha \tilde \alpha}L^{\alpha} \wedge (\sigma^{\tilde \alpha})^i_{j} \diff z^i \wedge \diff \bar z^{j} \ ,
\end{aligned}
\end{equation}
where $\sigma^{\tilde \alpha}$ are the standard Pauli matrices so that the two-forms $\iu (\sigma^{\tilde \alpha})^i_{j} \diff z^i \wedge \diff \bar z^{j}$ are anti-self-dual on the four-torus.

In the remainder of this work we want to analyze \eqref{eq:G4_susy_T4_f0} and, after giving its relation to type IIB flux compactifications, reinterpret it in terms of a solution of five-dimensional supergravity which arises from compactifying on $T^6$. The term in \eqref{eq:G4_susy_T4_f0} proportional to $f_0$ plays a special role from this point of view, as it contains a flux piece residing completely on $T^4$. This piece gives rise to a gauging of the axion dual to $C_3$ in the five-dimensional theory, making it a gauged supergravity. On the other hand, the four-form flux on $M_4$ will give a timelike profile for the axion as well, together with some self-dual fluxes. These self-dual fluxes arise in the almost-BPS solutions of \cite{Goldstein:2008fq,Bena:2009ev}, but here they combine with the gauging and the axion profile into a BPS configuration.

In the following we restrict to ungauged supergravity in five dimensions: we set $f_0 = 0$ and study the remaining fluxes. It might however be interesting to further consider the effect of gaugings on the supersymmetry of a solution. Furthermore, we will for simplicity  redefine $\bar c^a$ to absorb $C$ so that \eqref{eq:G4_susy_T4_f0} becomes
\begin{equation}
\label{eq:G4_susy_T4}
\begin{aligned}
 G_4^{\rm BPS} = & \Re(\bar c^a J_a\wedge \diff z^1 \wedge \diff z^2 )  + \iu f^{\alpha \tilde \alpha}L^{\alpha} \wedge (\sigma^{\tilde \alpha})^i_{j} \diff z^i \wedge \diff \bar z^{j} \ .
\end{aligned}
\end{equation}

\subsubsection{Interpretation as a type IIB compactification}

The above solutions can be straightforwardly dualized into type IIB flux solutions.
We write the four-torus as a product of a two-torus and two circles $T^4 = T^2\times S_1 \times S_2$. If we make both circles very small, we can perform a dimensional reduction on $S_1$ to weakly coupled type IIA and T-dualize along $S_2$ to IIB. If we take the decompactification limit for $S_2$, the geometry can be interpreted as a flux compactification on $M_4 \times T^2$ to four-dimensional flat spacetime. The spacetime-filling M2-branes become spacetime-filling D3-branes in this chain of dualities, and the solution belongs to the class of solutions described by  \cite{Grana:2000jj,Gubser:2000vg,Grana:2001xn} (see also \cite{Giddings:2001yu}). These solutions are type IIB Calabi-Yau compactifications with a constant dilaton, and the three-form flux $G_3 = F_3 - \tau H_3$ must be imaginary self dual (ISD) in order to fulfill the equations of motion. This implies that $G_3$ consists of a primitive $(2,1)$ piece, a $(0,3)$ piece and a piece that is equal to a $(0,1)$-form wedge $J$. In order for the vacuum to be supersymmetric, $G_3$ must be primitive $(2,1)$. Supersymmetry can be broken if $G_3$ has a $(0,3)$ component or a component which is a $(0,1)$-form wedge $J$.\footnote{Note that on a proper Calabi-Yau there are no harmonic one- or five-forms and primitivity is granted for any harmonic three-form.} Note that for this supersymmetry-breaking flux $G_3$ the equations of motion are still satisfied.
Dualizing the four-form in \eqref{eq:G4_susy_T4} gives a three-form flux $G_3$
\begin{equation}\label{ex:general_ISD}
 G_3 = c_a J^a \wedge \diff \bar z + F_-  \wedge \diff z \ ,
\end{equation}
where we set $F_- = f^{\alpha 1}L^{\alpha} + \iu f^{\alpha 2}L^{\alpha}$ and we ensured four-dimensional Lorentz invariance by setting $f^{\alpha 3}=0$.

The above solution breaks the supersymmetry of type IIB supergravity as follows: The hyper-K\"ahler-times-torus background and the presence of D3-branes each break half of supersymmetry so that we have $8$ supercharges in the theory, before considering fluxes. The gravitino variation with respect to these supercharges gives a mass matrix $P_{ij} = P^a \sigma^a_{ij}$ for the two corresponding gravitini in terms of the so-called $N=2$ prepotentials $P^a$, which are parameterized here by the flux $G_3$.\footnote{Here, $\sigma^a_{ij}$ are the symmetrized Pauli matrices, which are obtained from the standard Pauli matrices via $\sigma^a_{ij}= \varepsilon_{ik}(\sigma^a)^k_{j}$. }
These prepotentials can be understood as the $N=2$ generalization of the superpotential in an $N=1$ background, which is given in type IIB Calabi-Yau compactifications with D3-branes or O3-planes by the Gukov-Vafa-Witten (GVW) superpotential \cite{Gukov:1999ya}
\begin{equation}
 W = \int G_3 \wedge \Omega \ .
\end{equation}
If $G_3$ has a $(0,3)$ component, this induces a contribution for the F-term of the Calabi-Yau K\"ahler moduli.
The natural $N=2$ generalization of the GVW superpotential for hyper-K\"ahler-times-torus backgrounds is
\begin{equation}
 P^a = \int G_3 \wedge J^a \wedge \diff z \ .
\end{equation}
This gives us a first handle on the amount of supersymmetry preserved by the flux background.
An $N=2$ vacuum can only arise if all $P^a$ and their first derivatives with respect to the moduli vanish. This means that $G_3$ preserves eight supercharges if $G_3$ is of the form given in \eqref{ex:general_ISD}, but with $c_a=0$.
$N=2 \to N=1$ supersymmetry breaking can only arise if $P^a P^a = 0$ \cite{Louis:2009xd}.\footnote{For more details on $N=2 \to N=1$ breaking see \cite{Louis:2009xd,Louis:2010ui}.} This forces $G_3$ to be of the form \eqref{ex:general_ISD} with $c^a c^a = 0$, cf. \eqref{eq:condition_ca}. We conclude that the anti-self-dual flux component $F_-$ on $M_4$ does not break any supercharges, while the self-dual flux component breaks half of the supercharges if $c^a c^a=0$ holds and all supercharges otherwise.

%%%%%%%%%%%%%%%%%%%%%%%%%%%%%%%%%%%%%%%%%%%%%%%%%%%%%%%%%%%%%%%%%%%%%%%%%%%%%%%%%%%%
\subsection{BPS and almost-BPS black hole geometries}
\label{ss:BlackHole}

We give a short review of five-dimensional black hole, black ring and their microstate solutions, considered as $T^6$ compactifications of M-theory. We also discuss the relation to flux compactifications of the type given above.

The solutions we focus on fit inside an $N=2$ truncation of $N=8$ supergravity in five dimensions known as the STU model. The most general BPS solutions of this trunctation are known \cite{Gutowski:2004yv,Bena:2004de} and have the form
\eal{
\diff s_{11}^2 &=\; -(Z_1 Z_2 Z_3)^{-2/3}(\diff t + k)^2 + (Z_1 Z_2 Z_3)^{1/3} \,\diff s_4^2+(Z_1 Z_2 Z_3)^{1/3}\sum_{I=1}^3 \frac{\diff s_I^2}{Z_I}\ ,\\
G_4 &\equiv\; \diff A^{(I)}\wedge \omega_I = \sum_{I=1}^3 [- d \left(\frac{\diff t + k}{Z_I} \right)+ \Theta^{(I)} ]\wedge \omega_I\ ,\label{eq:ThreeChargeSol}
}
where $\diff s_I^2$ and $\omega_I$ are  unit metrics and unit volume forms on the three $T^2$'s inside $T^6$ and $ds_4^2$ is a four-dimensional hyper-K\"{a}hler metric. The one-form $k$ is supported on this four-dimensional base space and all functions appearing in the solution only depend on the base coordinates. The $\Theta^{(I)}$ are three two-forms on the hyper-K\"ahler base.  Note that the ansatz for the gauge fields relates the warp functions $Z_I$ appearing in the metric to the electric potentials which couple to three types of M2-branes. The $\Theta^{(I)}$ are magnetic fields on the hyper-K\"{a}hler base coupling to M5-branes.

\subsubsection{Supersymmetric three-charge solutions}
For a supersymmetric solution, the equations of motion for the ansatz \eqref{eq:ThreeChargeSol} reduce to
\eal{
\Theta^{(I)} &=\; - \star_4 \Theta^{(I)}\ ,\\
\Delta_4 Z_I &= \tfrac 12C_{IJK} \star_4(\Theta^{(J)}\wedge \Theta^{(K)})\ ,\\
dk - \star_4 d k &= Z_I \Theta^{(I)}\ .\label{eq:ThreeChargeSUSY}
}
The three two-forms $\Theta^{(I)}$ are anti-self-dual and they determine the warp factor and the rotation vector $k$.  Any solution to these equations is a 1/8 BPS solution of M-theory and a 1/2 BPS solution in $N=2$ supergravity in five dimensions. When the hyper-K\"ahler base is of Gibbons-Hawking form, one can find a solution in closed form in terms of 8 harmonic functions \cite{Gauntlett:2004qy, Bena:2005ni}, corresponding to 8 charges: the three M2 branes, three M5 brane dipole charges, and two geometric charges -- the Kaluza-Klein monopole and gravitational wave charges of the Gibbons-Hawking base. In this class we find black holes, black rings and their microstate geometries \cite{Bena:2007kg}, and these solutions when compactified to four dimensions descend to multicenter BPS black holes \cite{Bates:2003vx}.

\subsubsection{`Almost-BPS' three-charge solutions}

There exist non-BPS, extremal three-charge solutions, for which the metric and four-form still fit in the ansatz \eqref{eq:ThreeChargeSol}.\footnote{These are extremal in the sense that the asymptotic charges are those of an extremal, non-BPS black hole or black ring.} Because of there similarities to the BPS solutions discussed earlier, these are dubbed `almost-BPS' solutions \cite{Goldstein:2008fq,Bena:2009ev}.  These have  equations of motion that are formally very similar to the BPS ones.  In the language of \cite{Bena:2009fi}, we have
\eal{
\Theta^{(I)} &=\; + \star_4 \Theta^{(I)}\\
\Delta_4 Z_I &= \tfrac 12C_{IJK} \star_4(\Theta^{(J)}\wedge \Theta^{(K)})\;\\
dk + \star_4 d k &= Z_I \Theta^{(I)};\label{eq:ThreeChargeAlmostBPS}
}
We see that the magnetic fluxes $\Theta^{(I)}$ on the hyper-K\"ahler base space are self-dual for these solutions. When the hyper-K\"ahler base is Gibbons-Hawking, one can again construct explicit solutions. However, one cannot find a general solution in closed form in terms of only harmonic function, and the solutions are more messy than the BPS ones. See for instance \cite{Bena:2009ev,Bena:2009en,Bossard:2012ge}.

\subsubsection{Interplay with flux compactifications}
\label{sss:Interplay}

The equations of motion have similarities to those for a flux compactification of the type discussed above. As for those solutions, the magnetic fluxes obey a self-duality condition, and their square gives the Laplacian of the warp factors. In particular, we can interpret this solution as a flux compactification in the Becker-Becker class \cite{Becker:1996gj} when the solution has only one electric charge ($Z_1 \neq 1$) and we take
\be
Z_2=Z_3=1\ ,\qquad \Theta^{(1)}=0\ ,\qquad k=0\,,
\ee
and we make the identification
\be
\Theta^{(2)} = \Theta^{(3)} \equiv \Theta_\pm\ .
\ee
where $\pm$ denotes the Hodge duality eigenvalue on the hyper-K\"ahler space ($-$ for BPS solutions, $+$ for almost-BPS solutions).
The metric can be interpreted as a `compactification' to three dimensions:
\be
\diff s_{11}^2 =\; -Z_1^{-2/3}(\diff t^2 + \diff x^2 + \diff y^2) + Z_1^{1/3} \,(\diff s_4^2+ \diff \tilde s^2_4)\ ,\\ 	
\ee
with $x,y$ coordinates on the first $T^2$ inside $T^6$, $\diff \tilde s^2_4$ a unit metric on the complementary $T^4$.

\paragraph{BPS solutions.}

For the BPS solutions \eqref{eq:ThreeChargeAlmostBPS}, the four-form flux is $(2,2)$ and primitive and hence of the form \protect {\eqref{eq:G4_susy_T4}}. Comparing to \eqref{eq:G4_susy_T4}, the first term is absent ($\bar c_a = 0$) and the four-form is the wedge product of a $(1,1)$-form on $M_4$ and another (1,1) form on $T^4$:
\be
\Theta_-\wedge (\omega_2 + \omega_3) =  \iu f^{\alpha \tilde \alpha}L^{\alpha} \wedge (\sigma^{\tilde \alpha})^i_{j} \diff z^i \wedge \diff \bar z^{j}\,.
\ee

\paragraph{Almost-BPS solutions.}

The four-form flux of the almost BPS-solutions is a supersymmetry-breaking flux also in $N=8$. Comparing to Eq.\ \eqref{eq:G4_susy_T4}, only the first term is present. We can write the supersymmetry breaking flux as
\be
\Theta_+ \wedge (\omega_2 + \omega_3) =  c^a  J_a  \wedge \Re (\diff z^1 \wedge \diff z^2)  \ ,
\ee
where now $c^a$ are real coefficients. Note that the supersymmetry condition \eqref{eq:condition_ca} in particular says $\sum (c^a)^2 = 0$, which for real $c^a$ implies $c^a=0$. Thus, the class of almost-BPS solutions breaks all $32$ supercharges, as expected.

This gives a different view-point on almost-BPS solutions. So far they have been viewed as coming from flipping some signs in the three-charge solution. From the flux compactification perspective we see that all such self-dual fluxes fulfill the equations of motion, while supersymmetry imposes the additional condition \eqref{eq:condition_ca} which is not fulfilled for almost-BPS solution (in contrast to solutions with anti-self-dual fluxes, that are always supersymmetric).

\paragraph{Summary.}

We give an overview of the relation of BPS and almost-BPS solutions in the STU model to flux compactifications in Figure \ref{fig:Comparison}. Note that we do not have an interpretation of supersymmetric solution with a four-form flux \eqref{eq:G4_susy_T4} with the self-dual two-form $\bar c_a J^a\neq 0$.\footnote{Note that supersymmetry requires $\sum_a c_a^2 =0$} In the following section, we explore such flux solutions and how to interpret them as black-hole solutions.

%%%%%%%%%%%%%%%%%%%%%%%%%%%%%%%%%%%%%%%%%%%%%%%%%%%%%%%%%%%%%%%%%%%%%%%%%%%%%%%%%%%%
%%%%%%%%%%%%%%%%%%%%%%%%%%%%%%%%%%%%%%%%%%%%%%%%%%%%%%%%%%%%%%%%%%%%%%%%%%%%%%%%%%%%
\section{New solutions}
\label{s:Examples}

\begin{figure}[ht!]
\begin{center}
\includegraphics[width=.65\textwidth]{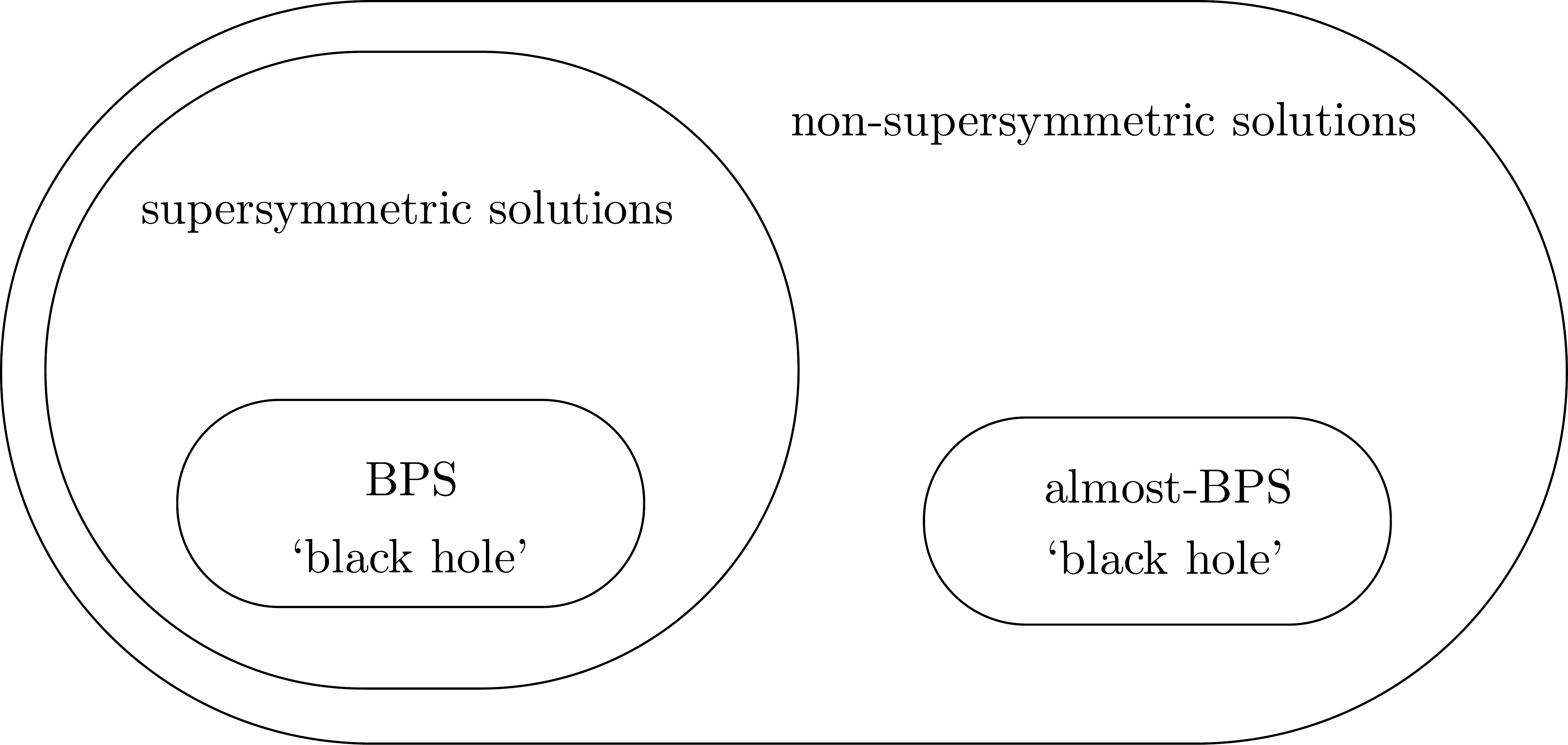}
\end{center}
\caption{Summary of the mapping of flux compactifications to black hole-type solutions. Any combination of $G_4$ components from the ``SUSY'' side results in a supersymmetric background. We also list the relation to the black hole BPS and almost-BPS solutions discussed in section \protect\ref{ss:BlackHole}. The flux \protect \eqref{eq:G4_Ansatz} can give solutions of any type in this diagram. This suggests that there are (interesting) flux compactification solutions, that we do not understand from the black hole side. We give the most general solution of this type in section \protect\ref{s:Examples}.\label{fig:Comparison}}
\end{figure}

In this section we want to treat in detail the class of supersymmetric flux vacua of the previous section that do not correspond to known classes of black hole solutions of the STU model.
We have seen that the BPS solutions of the STU model have real, anti-self-dual fluxes on the hyper-K\"ahler space, and the almost-BPS ones have real, self-dual ones that break the supersymmetry in eleven dimensions as well. We investigate the properties of the \emph{supersymmetric} flux solutions in the class \eqref{eq:G4_susy_T4} that have both self-dual and anti-symmetric fluxes, and we discuss how they fit inside an $N=2$ truncation of five-dimensional supergravity. We will see that these solutions break supersymmetry in the $N=2$ truncation, even though they are supersymmetric in eleven-dimensions and as $N=8$ supergravity solutions in five dimensions. See Figure \ref{fig:Comparison}.

%--------FIGURE: almostBPS,BPS,``new''BPS flux compactifications--
%-------end FIGURE------------------------------------------------

\subsection{The eleven-dimensional solution}

This solution is an extension of the solution described in the authors' recent work \cite{Bena:2011pi}. We consider the metric and four-form:
\begin{equation} \label{eq:11d_CY4} \begin{aligned}
\diff s_{11}^2 \;=\; & \e^{-2A}(-\diff t^2 + \diff x_{9}^2 + \diff x_{10}^2) + \e^{A} \diff s^2(X_8) \ ,\\
F_4 \; =\; &  \diff (e^{-3A} \operatorname{vol}_3) +\Im\,[(\Theta_+ - \iu\tilde \Theta_+) \wedge \diff z_1 \wedge \diff z_2 +(\Theta_- - \iu \ti\Theta_-)\wedge\diff z_1 \wedge \diff \bar z_2]\\
&+ \frac \iu 2 \bar \Theta_- (\diff z_1 \wedge \diff \bar z_1 - \diff z_2 \wedge \diff \bar z_2) \ ,
\end{aligned}
\end{equation}
Again, the subscripts denote the (anti-)self-duality properties. The anti-self-dual contributions do not break any supersymmetry. The self-dual pieces might break supersymmetry completely unless they obey an extra condition. Following the reasoning of the previous section (in particular considering \eqref{eq:condition_ca} and \eqref{eq:G4_susy_T4}), this gives the condition:
\be
(\Theta_+ - \iu\tilde \Theta_+)\wedge (\Theta_+ - \iu\tilde \Theta_+) =0\ .\label{eq:hol_2form}
\ee
Hence, this solution describes a supersymmetric flux compactification in the class \cite{Becker:1996gj}. When we keep one anti-self-dual flux $\Theta_-$ and put the other fluxes on the hyper-K\"ahler space to zero, the solution is a one-charge BPS solution as in section \ref{sss:Interplay}. If one tried to keep instead one self-dual flux, to obtain a one-charge almost-BPS solution (of the type discussed in section \ref{sss:Interplay}) the condition \eqref{eq:hol_2form} would then force this flux to be zero. Hence, an almost-BPS solution cannot be supersymmetric even if it only has one electric charge. 

We can make the supersymmetry more visible, through the specific projections to the preserved Killing spinors. Take coordinates $y^{1}\ldots y^4$ on $M_4$ and $y^5 \ldots y^8$ on $T^4$. The hyper-K\"ahler background breaks half of the supersymmetry, as it admits only a covariant spinor of (let us say) positive chirality. This corresponds to the projection $\Gamma^{1234} \eta = - \eta$. Furthermore, the flux $F_4$ might break more supersymmetry. Its electric component (corresponding to an M2-brane charge along the $9,10$ directions) breaks another half of supersymmetry, by the projection $\Gamma^{12345678} \eta = \eta$. The internal components of the flux can break additional supersymmetries. We prove there is one additional projector. Following \cite{Becker:1996gj}, the internal components of the Killing spinors obey $\slashed F \eta =0$ and $\slashed F_m \eta = 0$. The first condition gives:
\bea
0&=\tfrac 1 {4!}F_{ijkl} \Gamma^{ijkl} \eta \;=\;& \tfrac14[(\Theta_+)_{ij}\, \Gamma^{ij58} +(\tilde\Theta_+)_{ij}\,\Gamma^{ij68} ](1 - \Gamma^{5678})(1 -\Gamma^{1234})\eta\\
&&- \tfrac14[(\Theta_-)_{ij}\, \Gamma^{ij58} +(\tilde\Theta_-)_{ij}\,\Gamma^{ij68} +(\bar \Theta_-)_{ij}\,\Gamma^{ij56}](1 + \Gamma^{5678})(1 +\Gamma^{1234})\eta\ ,\nonumber
\eea
where we have inserted the projectors $(1 \pm \Gamma^{1234})/2$ by making use of the (anti-)self-duality of $\Theta_\mp$.

The terms containing the anti-self-dual components $\Theta_-,\ti \Theta_-$ vanish on the Killing spinors annihilated by the two two earlier projectors $1+\Gamma^{1234}$ and $1-\Gamma^{12345678} $, and this agrees with the known structure of BPS three-charge solutions, in which turning on an anti-self-dual field strength on the base does not affect the supersymmetry.

For arbitrary self-dual forms $\Theta_+, \ti \Theta_+$, the first line is not zero and supersymmetry is broken. However, for the specific choice
\eal{
\Theta_+ &=\; \theta_+ (e^1 \wedge e^3 + e^4 \wedge e^2)\ ,\\
\tilde \Theta_+ &=\; \theta_+(e^1 \wedge e^4 + e^2\wedge e^3) \ ,\label{eq:Theta_+}
}
this term contains a new projector:
\begin{equation}
0 =  \, 2 \theta_+ \Gamma^{1358}(1 + \Gamma^{3456})\eta \ ,
\end{equation}
which is compatible to the first two.  More generally, under the condition \eqref{eq:hol_2form} we always find such a projector and the solution has four supercharges.

It is not hard to see that the equations $\slashed F_m \eta = 0$ do not impose any extra conditions on the remaining Killing spinors, essentially because the flux pieces that are self-dual on the hyper-K\"ahler manifold always combine into the projector $1 + \Gamma^{3456}$, while the anti-self-dual components always give either $1 +\Gamma^{1234}$ or $1 + \Gamma^{5678}$, depending on the index $m$.
Therefore, the solution is $1/8$ BPS, and its 4 Killing spinors are annihilated by the projectors:
\begin{equation}
(1+\Gamma_{1234})\ ,\  (1 + \Gamma_{3456}) \ {\rm and }\ (1+\Gamma_{5678}) \ .\label{eq:SpinorsProjectionN=8}
\end{equation}

\subsection{Interpretation as \texorpdfstring{$N=2$}\~ truncation and supersymmetry}\label{ss:Interpretation}

We now analyze whether and how the solution \eqref{eq:11d_CY4} fits inside an $N=2$ truncation of five-dimensional supergravity.

Following the arguments of section \ref{ss:Truncations}, we know that our solution is naturally interpreted as a BPS solution of $N=8$ supergravity in five dimensions. When does it fit into an $N=2$ truncation of this theory? We discussed the condition in Table \ref{tab:Branching_SU6_vect}: we need to find a complex structure on the torus, such that the four-form legs on $T^6$ are of type (1,1). The solution \eqref{eq:11d_CY4} only has legs on $T^4$, so we restrict the problem to finding an appropriate complex structure on $T^4$. In general, we can always expand the four-form in the three self-dual and the three anti-self-dual two-forms on $T^4$.  Of these 6 components, a maximum of four can be turned into (1,1) forms by an appropriate choice of complex structure, while the remaining two form the holomorphic two-form. We now have two option:
\begin{itemize}
\item
Either we keep three anti-self-dual forms and one self-dual form. Because of the constraint \eqref{eq:hol_2form}, the self-dual component  must be zero, and we are left with a BPS solution in the $N=2$ truncation of the type BPS black hole type \eqref{eq:ThreeChargeSUSY}. This solution has already been discussed before \cite{Gauntlett:2002nw,Gutowski:2004yv,Bena:2004de}.
\item
The other possibility is that we keep three self-dual forms and one anti-self-dual form. In fact, in our solution \eqref{eq:11d_CY4} only two self-dual two forms on $T^4$ are allowed because of the demand of supersymmetry in $N=8$. Keeping one anti-self-dual two-form comes down to choosing two combinations out of the triple $\Theta_-,\tilde \Theta_-,\bar \Theta_-$ to be zero. These are the solutions with camouflaged supersymmetry.
\end{itemize}

Let us discuss the new second possibility. We choose $\Theta_-$ to be the only non-zero anti-self-dual two-form, such that $\tilde \Theta_-=\bar \Theta_-= 0$. If we take a complex structure such that the holomorphic one-forms on $T^4$ are
\be
\diff z \equiv \diff y ^8 + {\rm i} \diff y^5\,, \qquad \diff w \equiv \diff y^6 + {\rm i} \diff y^7\,,
\ee
then the four-form flux components on $T^6$ are all of type $(1,1)$:
\eal{
F_4 \; =\; &  \tfrac 12 \Im\,[(\Theta_+ - \iu\tilde \Theta_+) \wedge (\diff z \wedge \diff \bar z - \diff w \wedge \diff \bar w) - \Theta_- \wedge(\diff z \wedge \diff \bar z + \diff w \wedge \diff \bar w)]\\
&+\diff (e^{-3A} \operatorname{vol}_3) \ .\label{eq:FourForm2}
}
By the arguments of Section \ref{ss:Truncations}, the solution then fits in an $N=2$ truncation of five-dimensional $N=8$ supergravity.

The question remains if the solution preserves any of the supersymmetries in this truncation. Even though the solution is $1/8$ BPS in $N=8$ supergravity, it is not guaranteed that the truncation to $N=2$ keeps the unbroken supersymmetries. The projection conditions on the spinor for an $1/8$ BPS solution of $N=8$ were given in equation \eqref{eq:SpinorsProjectionN=8}. In Section \ref{ss:SpinorTruncations} we also discussed the projection conditions on the eleven-dimensional spinor following from the $N=2$ truncation. In particular, the first condition in \eqref{eq:SpinorsProjection}:
\be
 (1 - \Gamma_{3456})\, \eta = 0\ ,
\ee
is incompatible with the second condition in \eqref{eq:SpinorsProjectionN=8}. We find that the unbroken supersymmetries are projected out by the $N=2$ truncation: the four-form flux \eqref{eq:FourForm2} gives a solution that is  supersymmetric in $N=8$, but cannot be a supersymmetric solution of any $N=2$ truncation.

%%%%%%%%%%%%%%%%%%%%%%%%%%%%%%%%%%%%%%%%%%%%%%%%%%%%%%%%%%%%%%%%%%%%%%%%%%%%%%%%%%%%
%%%%%%%%%%%%%%%%%%%%%%%%%%%%%%%%%%%%%%%%%%%%%%%%%%%%%%%%%%%%%%%%%%%%%%%%%%%%%%%%%%%%
\section{Single-center example}\label{s:Explicit}

We discuss a specific solution of the form \eqref{eq:11d_CY4} with one source. We choose a solution as in the previous subsection, whose supersymmetry gets camouflaged in any $N=2$ truncation, with non-vanishing fluxes $\Theta_+,\tilde \Theta_+,\Theta_-$. We choose the HK metric to be Taub-NUT:
\be
ds_4^2 = V^{-1} (\sigma_3)^2 + V (dr^2 + r^2 ((\sigma_1)^2 + (\sigma_2)^2)\,,
\ee
where $V$ is a harmonic function
\be
V = h + \frac{q}{r}\,,
\ee
and $\sigma_i$ are right-invariant one-forms on $\SU(2)$:
\eal{
\sigma_1 &= \sin \psi \,\diff \theta - \cos \psi \sin \theta \,\diff \phi\,,\\
\sigma_2 &= \cos \psi \,\diff \theta + \sin \psi \sin \theta \,\diff \phi\,,\\
\sigma_3 &=  \diff \psi + \cos \theta \,\diff \phi\,.
}

We take the vielbeins
\be
e^1 = \sqrt V \diff r\,,\quad e^2 = r\sqrt V \sigma_1\,,\quad e^3 = r\sqrt V \sigma_2\,, \quad e^4 = \frac 1 {\sqrt V} \sigma_3\, ,
\ee
The self-dual fluxes are as in eq.\ \eqref{eq:Theta_+}:
\eal{
\Theta_+ &= r \,\theta_+(r) (V dr \wedge \sigma^1 + \sigma_2 \wedge \sigma_3) \,, \\
\tilde \Theta_+ &= r\, \theta_+(r) (V dr \wedge \sigma^2 + \sigma_3 \wedge \sigma_1)\,.
}
For $\Theta_+,\tilde \Theta_+$ to be closed, we must have 
\be
\theta_+ = \frac {k_+} r\,,
\ee
with $k_+$ a constant.

The anti-self-dual flux is (see for instance \cite{Bena:2007kg})
\be
\Theta_- = \sum_{a=1}^3\partial_a \left(\frac{K}V\right) \Omega_-^{(a)}\,,
\ee
where $\Omega^{(a)}_-$ a basis of anti-self-dual fluxes
\be
\Omega^{(a)}_- = e^4 \wedge e^{a} -\tfrac 12 \epsilon_{abc} e^b \wedge e^c\,,\qquad  a,b,c = 1,2,3\, ,
\ee
and $K$ is an arbitrary harmonic function. Since we focus on a single-center solution, we choose it to be 
\be
K = \frac{k_-}{r}\,.
\ee
The equation of motion for the warp factor $Z \equiv e^{3A}$ is:
\be
\Delta_4 Z = \Theta^2_+ +\tilde \Theta^2_+ + \Theta_-^2
\ee
With our choice of fluxes, this becomes
\be
\Box_3 Z = 2 V \Box_3 \theta_+^2 + \Box_3 \frac{K^2}{V}\,,
\ee
where $\Box_3$ is the Laplacian on the three-dimensional flat base. The solution is given by:
\be
Z = L + \frac{K^2}{V} + 2\frac{h k_+^2}{r^2} + \frac 23 \frac{q k_+^2}{r^3} \,,
\ee
where  $L$ is another harmonic function describing an M2-brane source:
\be
L = 1 + \frac{q_{M2}}{r}\,.
\ee
Note that this warp factor now contains a mixture of BPS and almost-BPS looking terms.

We discuss the physics of the solution. Asymptotically, the solution carries only charge of M2-branes wrapped on directions $x_1,x_2$:
\be
Q_{M2} = q_{M2} + \frac{k_-^2}{q}\,.
\ee
We investigate the near-horizon region $r\to 0$. After a rescaling of the time coordinate, it can be brought into the form of an $AdS_4 \times S^3$ times a $T^4$ factor whose size runs with the radial coordinate of $AdS_4$:
\be
ds_{11}^2 = \frac{r^2}{R^2}(-\diff t^2+\diff x_1^2 + \diff x_2^2) + R^2 \frac{\diff r^2}{r^2} + \frac{R^2}{q^2} ds^2 (S^3)+\frac {R^2}{q r} ds^2 (T^4)
\ee
with
\be
R = \left(\tfrac23 q^4 k^2\right)^{1/6}\,.
\ee

One can alternatively interpret this solution in five dimensions, where the near horizon geometry is 
$AdS_2 \times S^3$ and the metric is that of a black hole with a finite horizon area $A_H \propto \frac{R^3}{q^3}$. However, unlike usual three-charge black hole solutions, this solution is singular because the scalars blow up at the horizon: the two-torus with coordinates $x_1,x_2$ shrinks to zero size at the horizon, while the $T^4$ volume blows up. This is expected from the fact that this solution only has one charge, and there is no five-dimensional BPS black hole with only one charge. It would be interesting to further explore the solutions of the type (\ref{eq:11d_CY4}) to see what other solutions one can construct in this class.

%%%%%%%%%%%%%%%%%%%%%%%%%%%%%%%%%%%%%%%%%%%%%%%%%%%%%%%%%%%%%%%%%%%%%%%%%%%%%%%%%%%%
%%%%%%%%%%%%%%%%%%%%%%%%%%%%%%%%%%%%%%%%%%%%%%%%%%%%%%%%%%%%%%%%%%%%%%%%%%%%%%%%%%%%
\section*{Acknowledgments}

We would like to thank G.\ Bossard, S.\ Ferrara, S.\ Giusto, M.\ Gunaydin, A.\ Marrani and T.\ Van Riet for valuable discussions and especially M. Gra\~na for her collaboration during the early stages of this project. This work was supported in part by the ANR grant 08-JCJC-0001-0 and by the ERC Starting Independent Researcher Grants 240210 -- String-QCD-BH and 259133 -- ObservableString.

\appendix{

\section{Hyper-K\"ahler geometry}
\label{app:HyperKahler}

A four-dimensional hyper-K\"ahler manifold has by definition the holonomy group $Sp(1) \equiv \SU(2)$, which is equivalent to the existence of o6ne covariantly constant spinor $\eta$, i.e. $\nabla \eta = 0$. This spinor $\eta$ and its charge conjugate $\eta^c$ define a triple of harmonic two-forms $J^a$, $a=1,2,3$, by\footnote{Note that charge conjugation in four Euclidean dimensions preserves chirality.}
\begin{equation}
J^1+\iu J^2 = \bar \eta^c \gamma_{mn} \eta \diff x^m \wedge \diff x^n \ , \qquad  J^3 = \iu \bar \eta \gamma_{mn} \eta \diff x^m \wedge \diff x^n \ .
\end{equation}
With the help of Fierz identities one can show that they obey
\begin{equation}
 J^a \wedge J^b = \delta^{ab} {\rm vol}_4 \ .
\end{equation}
In particular, all $J^a$ have positive norm.
The choice of these three harmonic two-forms is equivalent to the choice of $\eta$ and determines the geometry completely. Moreover, there is an $\SU(2)$ gauge freedom to rotate $\eta$ and $\eta^c$ into each other, which translates into $\SO(3)$ rotations of the $J^a$. In a heterotic (type II) string compactification this $\SU(2)$ forms (part of) the R-symmetry group.

In general, the space of harmonic two-forms on a hyper-K\"ahler manifold is of signature $(3,n)$, where the $J^a$ form the basis for a maximal subspace of positive signature.\footnote{If the manifold is compact, it is K3 and $n=19$.} If we denote an orthonormal base (with respect to the wedge product) of harmonic two-forms by $\{ J^a, L^\alpha \}$, $\alpha=1, \dots, n$,
the Hodge star operator $\star_4$ can be defined via
\begin{equation}
 \star_4 J^a = J^a \ , \qquad \star_4 L^\alpha = - L^\alpha \ .
\end{equation}
This means that the $J^a$ ($L^\alpha$) are (anti-)self-dual.
On the hyper-K\"ahler manifold exists a triplet of complex structures $I^a$, which are constructed from the hyper-K\"ahler two-forms $J^a$ by contraction with the inverse metric. If we choose one complex structure, let us say $I^3$, the $n+1$ two-forms $J^3$ and $L^\alpha$ are $(1,1)$ while $J^1 + \iu J^2$ defines a holomorphic two-form.

}

\bibliographystyle{toine}
\bibliography{FuzzComp}

% \begin{thebibliography}{10}
% \end{thebibliography}

\end{document}